\documentclass[aps,prd,preprint,preprintnumbers,unsortedaddress,superscriptaddress,showpacs,nofootinbib]{revtex4-1}

\usepackage{graphicx,color}
\usepackage{relsize}
\usepackage{slashed}
\usepackage{color}
\usepackage{ulem}
\usepackage{tabu}

\usepackage{amsmath}
\usepackage{amssymb}
\usepackage{amsthm}
\usepackage{mathrsfs}
\usepackage{graphicx}
\usepackage{epstopdf}
\usepackage{fancyhdr}
\usepackage{array}
\usepackage[all]{xy}
\usepackage{eufrak}
\usepackage{euscript}
\usepackage{enumerate}
\usepackage{slashed}
\usepackage{hyperref}
\usepackage{subfigure} 
\usepackage{epstopdf} 

\hypersetup{pdftex,colorlinks=true,linkcolor=blue,citecolor=blue,menucolor=black,urlcolor=blue,filecolor=blue}

\newcommand{\myvec}[1]{\ensuremath{\begin{pmatrix}#1\end{pmatrix}}}

\begin{document}

\title{Molecular $\Omega_c$ states generated from coupled meson-baryon channels}

\author{V.~R.~Debastiani}
\email{vinicius.rodrigues@ific.uv.es}
\affiliation{Departamento de F\'{i}sica Te\'{o}rica and IFIC, Centro Mixto Universidad de Valencia - CSIC,
Institutos de Investigaci\'{o}n de Paterna, Aptdo. 22085, 46071 Valencia, Spain.}

\author{J.~M.~Dias}
\email{jdias@if.usp.br}
\affiliation{Departamento de F\'{i}sica Te\'{o}rica and IFIC, Centro Mixto Universidad de Valencia - CSIC,
Institutos de Investigaci\'{o}n de Paterna, Aptdo. 22085, 46071 Valencia, Spain.}
\affiliation{Instituto de F\'{i}sica, Universidade de S\~{a}o Paulo, Rua do Mat\~{a}o, 1371, Butant\~{a}, CEP 05508-090, S\~{a}o Paulo, S\~{a}o Paulo, Brazil}

\author{W.~H.~Liang}
\email{liangwh@gxnu.edu.cn}
\affiliation{Department of Physics, Guangxi Normal University, Guilin 541004, China.}

\author{E.~Oset}
\email{eulogio.oset@ific.uv.es}
\affiliation{Departamento de F\'{i}sica Te\'{o}rica and IFIC, Centro Mixto Universidad de Valencia - CSIC,
Institutos de Investigaci\'{o}n de Paterna, Aptdo. 22085, 46071 Valencia, Spain.}

\begin{abstract}
We have investigated $\Omega_c$ states that are dynamically generated from the meson-baryon interaction. We use an extension of the local hidden gauge to obtain the interaction from the exchange of vector mesons. We show that the dominant terms come from the exchange of light vectors, where the heavy quarks are spectators. This has as a consequence that heavy quark symmetry is preserved for the dominant terms in the $(1/m_Q)$ counting, and also that the interaction in this case can be obtained from the $\textrm{SU(3)}$ chiral Lagrangians. We show that for a standard value for the cutoff regulating the loop, we obtain two states with $J^{P}={1/2}^{-}$ and two more with $J^{P}={3/2}^{-}$, three of them in remarkable agreement with three experimental states in mass and width. We also make predictions at higher energies for states of vector-baryon nature.
\end{abstract}



\maketitle

\section{Introduction}
\label{sec:intro}

In Ref.~\cite{lhcbexp} the LHCb collaboration reported five new narrow $\Omega^0_c$ states studying the $\Xi^+_c K^-$ mass spectrum produced in high energy $pp$ collisions: $\Omega_c(3000)$, $\Omega_c(3050)$, $\Omega_c(3066)$, $\Omega_c(3090)$ and $\Omega_c(3119)$. Predictions for such states and related ones had been done within quark model in Refs.~\cite{ebert,roberts,garcilazo,migura,ebert2,valcarce,shah,vijande,yoshida,cheng,cheng2,chiladze,chiralquark}. Molecular states had also been used to make predictions in Refs.~\cite{hofmann,tejero} studying the interaction of coupled channels, one of them the $\Xi^+_c K^-$ where the recent LHCb states were
found. A more updated study along these lines was done in Ref.~\cite{romanets}, where predictions for charmed and strange baryons are done using an interaction based on $\textrm{SU(6)}$ flavor-spin symmetry in the light quark sector and $\textrm{SU(2)}$ spin symmetry in the heavy quark sector, extending the $\textrm{SU(3)}$ Weinberg-Tomozawa interaction. All these works take the coupled channels of meson-baryon that couple to the desired baryon quantum numbers and use a unitary scheme to obtain the scattering matrix between the channels, looking for poles of this matrix. The differences come from the input interaction and the way that
loops are regularized.

The experimental findings of Ref.~\cite{lhcbexp} have brought a new wave of theoretical activity with many suggestions to explain the new states. Different versions of quark models have been proposed in Refs.~\cite{rosner,qzhao,weiwang,xiangliu}. Pentaquark options have been suggested
in Refs.~\cite{yang,huangping,polyakov,an,sheldon,anisovich}. QCD sum rules were used to describe these states in Refs.~\cite{azizi,chenhosaka,zhiwang,aliev,sundu,maohosaka,sundu2,Wang:2017xam}. Lattice QCD has also brought some light into the problem \cite{mathur}. Some works have emphasized the value of decay properties to obtain information on the nature of these states \cite{zezhao,wangzhao,kim} and a discussion on the possible quantum numbers was done in Ref.~\cite{chengchiang}.

In the molecular picture, an update of the work of Ref.~\cite{tejero} was done in Ref.~\cite{angels} using some information from the experimental spectrum to regularize the loops and then giving a
description of the mass and width of two states of Ref.~\cite{lhcbexp} as $J^{P}={1/2}^{-}$ meson-baryon molecular states.

In the present work we shall follow Refs.~\cite{romanets,angels} for the coupled channels and the unitarization procedure. We differ in the input for the interaction, which in our case is based on the local hidden gauge approach, exchanging vector mesons \cite{hidden1,hidden2,hidden4,Yama,hideko}.

We must clarify this concept. The local hidden gauge approach \cite{hidden1,hidden2,hidden4,Yama} works with pseudoscalar and vector mesons in the light sector and chiral symmetry is one of its assets, showing up in the limit of small mass of the pseudoscalar mesons (Goldstone bosons). In 
Refs.~\cite{hidden1,hidden2,hidden4,Yama,hideko,derafael}, and particularly in Refs.~\cite{Yama,derafael}, one can see that the terms of the chiral Lagrangians can be obtained from the exchange of vector mesons in the local hidden gauge. Ref.~\cite{derafael} also shows that the consideration of vector mesons is necessary to implement vector meson dominance. Both in Ref.~\cite{derafael} and Ref.~\cite{Yama} it is also shown that the formalisms using antisymmetric tensors for the vector mesons, and the use of ordinary vector fields in the local hidden gauge are equivalent. If one specifies to the meson-baryon Lagrangians \cite{ecker}, it is easy to show that the exchange of vector mesons gives rise exactly to the lowest order chiral Lagrangian in the limit of small momentum transfer compared to the vector meson mass. All this occurs within SU(3), involving $u$, $d$, $s$ quarks. The local hidden gauge in the unitary gauge in SU(3), can be found in Ref.~\cite{derafael} and with more detail in Ref.~\cite{hideko}. The extrapolation to SU(4) to incorporate $c$ quarks, or even higher with $b$ quarks, is not straightforward, as one cannot invoke the Goldstone boson character for $D$ or $B$ mesons.
Yet, what one does is the following: think of the $DN$ interaction for instance. In the $D^0 \, p \to D^0 \, p$ transition we have $c \bar u$ in the $D^0$ and $u u d$ quarks in the $p$, then we can only exchange $\rho^0$, $\omega$ vector mesons and the $c$ quark of the $D^0$ is a spectator. In this case the situation is the same as in $\bar{K}^0 \, p \to \bar{K}^0 \, p$. The $s$ quark of the $\bar{K}^0$ ($s \bar d \,$) is also a spectator and only $\rho$, $\omega$ vector mesons are exchanged.
In as much as the $c$ quark in $D^0 \, p \to D^0 \, p$ is a spectator, the dynamics is the same as in the $\bar{K}^0 \, p \to \bar{K}^0 \, p$ transition, and for this we can use the local hidden gauge approach. We find thus a way to obtain the $D^0 \, p \to D^0 \, p$ interaction using the dynamics of the light quark sector, since only these quarks are also involved in this case. Hence, in the diagonal channels the interaction is well controlled.

However, assume the coupled channel $\pi \, \Sigma_c$, then in the transition $D^0 \, p \to \pi^0 \, \Sigma_c^+$, if we extrapolate the local hidden gauge approach to SU(4), we would be exchanging a $D^\ast$ and the $c$ quarks are now involved. This is an extrapolation of the local hidden gauge approach which is model dependent.
Fortunately, the exchange of $D^\ast$ is penalized with respect to the exchange of light vector mesons by a factor of $\displaystyle \left(\frac{m_\rho}{m_{D^\ast}}\right)^2$, which is a small factor and then one is only introducing uncertainties in some non diagonal terms which are very small.
Formally one can use the SU(4) extrapolation of the local hidden gauge approach and for the diagonal terms the framework automatically filters the exchange of light vectors, providing the results that one obtains from the mapping explained before. This is what is done in Ref.~\cite{angels}.


In the present work the diagonal terms that we evaluate coincide with those of Ref.~\cite{angels} where the model of Ref.~\cite{hofmann} is used implementing also the exchange of vector mesons and $\textrm{SU(4)}$ symmetry for mesons and baryons.
We, instead, use explicit wave functions for the baryon states imposing flavor-spin symmetry on the light quark sector and singling out the heavy quarks.
Hence, in the baryon sector we are not using SU(4) symmetry.
For the diagonal terms we also show that one is exchanging light vectors and the heavy quarks are spectators.
In this case we obtain the same matrix elements as in Ref.~\cite{angels}, but there are differences in the non diagonal ones. Since in the dominant terms we are exchanging only light vectors and the heavy quarks are spectators, the interaction automatically respects heavy quark symmetry \cite{isgur,neubert,manohar}.
The non diagonal terms which exchange heavy vectors do not fulfill heavy quark symmetry, but neither should them since these are terms of order $O(m_Q^{-2})$ in the heavy quark mass counting.
In addition to the work of Ref.~\cite{angels} we also include pseudoscalar-baryon(${3/2}^+$) components and we obtain two more states. We can identify two states of $J^{P}={1/2}^{-}$ and one of $J^{P}={3/2}^{-}$ with the states found in Ref.~\cite{lhcbexp}. We also look for vector-baryon states and find three states at higher energies.

\section{Formalism}
\label{sec:form}

Following Ref.~\cite{romanets} we distinguish the cases with $J^{P}={1/2}^-$ and $J^{P}={3/2}^-$ and write the coupled channels. In Ref.~\cite{romanets} 12 coupled channels are used ranging from thresholds $2965$ MeV to $3655$ MeV. The experimental states of Ref.~\cite{lhcbexp} range from $3000$ MeV to about $3120$ MeV. Hence we restrict our space of meson-baryon states up to the $\Omega_c\, \omega$
with mass $3478$ MeV. Yet, the diagonal matrix element in this channel is zero and we can also eliminate it. The energy ranged by the
channels chosen widely covers the range of energies of Ref.~\cite{lhcbexp} and it is a sufficiently general basis of states. We show in Tables \ref{tab1} and
\ref{tab2} these states together with their threshold masses.

\begin{table}[h!]
\caption{$J=1/2$ states chosen and threshold mass in MeV.}
\centering
\begin{tabular}{c | c c c c c c c}
\hline\hline
{\bf States} ~& ~$\Xi_c\bar{K}$~ & ~$\Xi^{\prime}_c\bar{K}$~ & ~$\Xi D$~ & ~$\Omega_c \eta$ ~&~ $\Xi D^*$ ~& ~$\Xi_c \bar{K}^*$ ~& ~$\Xi^{\prime}_c\bar{K}^*$\\
\hline
{\bf Threshold} ~& $2965$ & $3074$ & $3185$ & $3243$ & $3327$ & $3363$ & $3472$\\
\hline\hline
\end{tabular}
\label{tab1}
\end{table}

\begin{table}[h!]
\caption{$J=3/2$ states chosen and threshold mass in MeV.}
\centering
\begin{tabular}{c | c c c c c c }
\hline\hline
{\bf States} ~& ~$\Xi^*_c\bar{K}$~ & ~$\Omega^*_c \eta$~ & ~$\Xi D^*$~ & ~$\Xi_c \bar{K}^*$ ~&~ $\Xi^* D$ ~& ~$\Xi^{\prime}_c \bar{K}^*$ ~\\
\hline
{\bf Threshold} ~& $3142$ & $3314$ & $3327$ & $3363$ & $3401$ & $3472$\\
\hline\hline
\end{tabular}
\label{tab2}
\end{table}

\begin{figure}
	\begin{center}
		\includegraphics[width=0.8\textwidth]{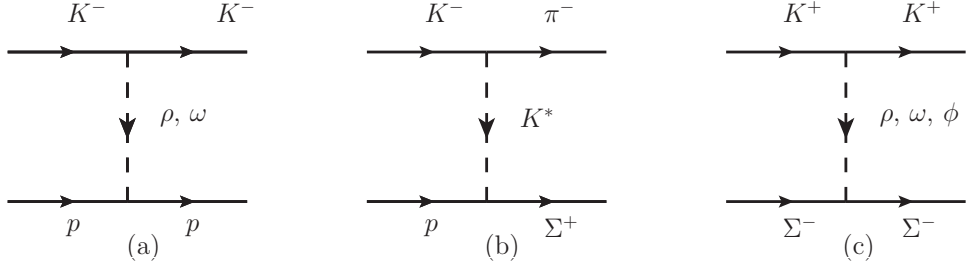}
	\end{center}
	\caption{\label{diagXic} Vector exchange in the meson-baryon interaction.}
\end{figure}

The meson-baryon interaction in the $\textrm{SU(3)}$ sector is given by the chiral Lagrangian \cite{ecker,mizutani}
\begin{eqnarray}
\label{lag}
\mathcal{L}^B = \frac{1}{4f_\pi^2}\, \langle \bar{B} i \gamma^{\mu} \Big[ (\Phi\, \partial_{\mu}\Phi -  \partial_{\mu}\Phi\,\Phi\,)B -
B(\Phi\, \partial_{\mu}\Phi -  \partial_{\mu}\Phi\,\Phi\,) \Big ] \rangle \, ,
\end{eqnarray}
where $\Phi$, $B$ are the $\textrm{SU(3)}$ matrices for pseudoscalar mesons and baryons
\begin{equation}
\label{phimatrix}
\Phi =
\left(
\begin{array}{ccc}
\frac{1}{\sqrt{2}} \pi^0 + \frac{1}{\sqrt{6}} \eta & \pi^+ & K^+ \\
\pi^- & - \frac{1}{\sqrt{2}} \pi^0 + \frac{1}{\sqrt{6}} \eta & K^0 \\
K^- & \bar{K}^0 & - \frac{2}{\sqrt{6}} \eta
\end{array}
\right)\, ,
\end{equation}

\begin{equation}
\label{Bmatrix}
B =
\left(
\begin{array}{ccc}
\frac{1}{\sqrt{2}} \Sigma^0 + \frac{1}{\sqrt{6}} \Lambda &
\Sigma^+ & p \\
\Sigma^- & - \frac{1}{\sqrt{2}} \Sigma^0 + \frac{1}{\sqrt{6}} \Lambda & n \\
\Xi^- & \Xi^0 & - \frac{2}{\sqrt{6}} \Lambda
\end{array}
\right) \, .
\end{equation}
The symbol $\langle \,\, \rangle$ stands for the $\textrm{SU(3)}$ trace and $f_\pi=93$ MeV is the pion
decay constant. At energies close to threshold one can consider only the dominant contribution coming
from $\partial_0$ and $\gamma^0$ \cite{OOR}, such that the interaction is given by
\begin{eqnarray}
\label{kernel}
V_{ij}= -C_{ij} \frac{1}{4f_\pi^2}(k^0+k^{\prime 0})\, ,
\end{eqnarray}
where $k^0$, $k^{\prime 0}$ are the energies of the incoming and outgoing mesons, respectively,
\begin{equation}\label{k0}
  k^0 = \frac{s + m_{m_i}^2 -M_{B_i}^2}{2\sqrt{s}}\, , \quad \quad k^{\prime 0} = \frac{s + m_{m_j}^2 -M_{B_j}^2}{2\sqrt{s}}\, \
\end{equation}
where $m_{m_i}$, $M_{B_i}$ ($m_{m_j}$, $M_{B_j}$) are the masses of the initial (final) meson, baryon, respectively,
 and
$C_{ij}$ are coefficients early calculated which are tabulated in Ref.~\cite{mizutani} for the case of $K^- p$ and coupled channels. The extension of Eq.~\eqref{kernel} to the charm sector is complicated particularly in the baryon sector. Yet, using the local hidden gauge approach \cite{hidden1,hidden2,hidden4,Yama,hideko} the task is notably simplified and clarified simultaneously. In the hidden gauge approach the meson-baryon interaction in $\textrm{SU(3)}$ is obtained exchanging vector mesons as in Fig.~\ref{diagXic}.

The ingredients needed are the vector(V)-pseudoscalar(P)-pseudoscalar(P) Lagrangian
\begin{eqnarray}
\label{vppLag}
\mathcal{L}_{VPP} = -i g\, \langle \,\, [\,\Phi,\,\partial_{\mu} \Phi\,]\, V^{\mu}\, \rangle \, ,
\end{eqnarray}
with
\begin{equation}
\label{vfields}
V_\mu=\left(
\begin{array}{ccc}
\frac{1}{\sqrt{2}}\rho^0+\frac{1}{\sqrt{2}}\omega & \rho^+ & K^{*+}\\
\rho^- &-\frac{1}{\sqrt{2}}\rho^0+\frac{1}{\sqrt{2}}\omega & K^{*0}\\
K^{*-} & \bar{K}^{*0} &\phi\\
\end{array}
\right)_\mu\ ,
\end{equation}
and the vector(V)-baryon(B)-baryon(B) Lagrangian
\begin{eqnarray}
\label{vbbLag}
\mathcal{L}_{VBB} = g\, \Big( \langle \bar{B} \gamma_{\mu} [V^{\mu},B] \,\rangle +
\langle \bar{B} \gamma_{\mu} B \rangle \langle V^{\mu} \rangle \Big) \, ,
\end{eqnarray}
with $g=m_V/2f_\pi$ and $m_V$ the mass of the vector mesons (we take an average of about
800 MeV).

It is easy to prove that the picture of the vector meson exchange with these Lagrangians gives
rise to the same interaction as Eq.~\eqref{kernel} taking
$q^2/m^2_V\to 0$ in the propagator of the exchanged vector, which is quite good at low energies.
One can even keep this term in the propagator, as done in Ref.~\cite{xiaooz}, since in the meson-meson
sector this is shown to generate higher order terms of the Lagrangian 
\cite{derafael,Yama}. Yet, if one takes a regulator of the loops integrating to a value $|\vec{q}_{max}|$
and fitting this to data, the consideration of the $q^2/m^2_V$ terms in the vector propagator is
unnecessary.

Extending Eqs.~\eqref{vppLag} and \eqref{vbbLag} to the charm sector is easy for the
$VPP$ Lagrangian but not for the $VBB$ Lagrangian \cite{hofmann}, but we introduce here a procedure
that renders it very easy. For this, let us look at the quark structure of the $\rho^0$, $\omega$ and $\phi$
(which can be extended to $K^*$, $\rho^{\pm}$)
\begin{eqnarray}
\label{waves}
\rho^0 &=& \frac{1}{\sqrt{2}} (u\bar{u} - d\bar{d})\, ,\nonumber \\
\omega &=& \frac{1}{\sqrt{2}} (u\bar{u} + d\bar{d})\, ,\nonumber \\
\phi &=& s\bar{s} \, .
\end{eqnarray}
In the approximation of taking $\gamma^{\mu} \to \gamma^0$ the spin dependence disappears, and we can consider an operator at the quark level as in Eq.~\eqref{waves}. We can take for instance
\begin{eqnarray}
\label{wavep}
\langle p |\,g\,\rho^0\,| p \rangle \equiv \frac{1}{\sqrt{2}}\frac{1}{\sqrt{2}}
\langle \phi_{MS}\, \chi_{MS} + \phi_{MA}\, \chi_{MA} | g \frac{1}{\sqrt{2}} (u\bar{u} - d\bar{d}) |
\phi_{MS}\, \chi_{MS} + \phi_{MA}\, \chi_{MA} \rangle \, , \nonumber\\
\end{eqnarray}
where $\phi_{MS}$, $\phi_{MA}$, $\chi_{MS}$, $\chi_{MA}$ are the flavor and spin mixed symmetric and mixed antisymmetric wave functions for the proton \cite{close}. Then, we can see that one gets the same result as using Eq.~\eqref{vbbLag}, and this is also the case for all transitions. Therefore we use this method to obtain the $VBB$ vertex in the charm sector. The extension of the vertex $\mathcal{L}_{VPP}$ to the charm sector is easier. We take the same structure as in Eq.~\eqref{vppLag} but now $P$ and $V$ are
\begin{equation}
\label{Pmatrix}
P =
\left(
\begin{array}{cccc}
\frac{1}{\sqrt{2}} \pi^0 + \frac{1}{\sqrt{3}} \eta  + \frac{1}{\sqrt{6}} \eta^{\prime} & \pi^+ & K^+ & \bar{D}^0 \\
\pi^- &  - \frac{1}{\sqrt{2}} \pi^0 +\frac{1}{\sqrt{3}} \eta + \frac{1}{\sqrt{6}} \eta^{\prime} & K^0 & D^- \\
K^- & \bar{K}^0 & -\frac{1}{\sqrt{3}} \eta +\sqrt{\frac{2}{3}} \eta^{\prime} & D_s^- \\
D^0 & D^+ & D_s^+ & \eta_c\\
\end{array}
\right)\, ,
\end{equation}
where we include the mixing between $\eta$ and $\eta^{\prime}$ \cite{bramon}, and

\begin{equation}
\label{Vmatrix}
V =
\left(
\begin{array}{cccc}
\frac{1}{\sqrt{2}} \rho^0 + \frac{1}{\sqrt{2}} \omega & \rho^+ & K^{* +} & \bar{D}^{* 0} \\
\rho^- & -\frac{1}{\sqrt{2}} \rho^0 + \frac{1}{\sqrt{2}} \omega & K^{* 0} & \bar{D}^{* -} \\
K^{* -} & \bar{K}^{* 0} & \phi & D_s^{* -} \\
D^{* 0} & D^{* +} & D_s^{* +} & J/\psi\\
\end{array}
\right)\, .
\end{equation}

It has been shown in Ref.~\cite{luisbd} (see section IIA of that reference), using similar arguments at the quark level as in Eq.~\eqref{wavep}, that in the heavy sector the coupling of the light vectors to the charmed mesons leaves the heavy quark as a spectator. Then one can map the matrix elements with light quarks to the equivalent ones in $\textrm{SU(3)}$, with the result that using Eq.~\eqref{vppLag} in $\textrm{SU(4)}$, with the matrices of Eqs.~\eqref{Pmatrix} and \eqref{Vmatrix}, the result obtained is the same as using this quark model with the heavy quarks as spectators. In other words, one is making use of the $\textrm{SU(3)}$ content of $\textrm{SU(4)}$. Furthermore, the fact that the heavy quarks are spectators has immediately as a consequence that the interaction complies with the rules of heavy quark spin symmetry (HQSS). However, if we have non diagonal transitions like $\Xi_c \bar{K}\to \Xi D$ one must exchange a $D^*_s$ and the heavy quarks are involved. Here $\textrm{SU(4)}$ is used and the result is more model dependent, apart from not satisfying the rules of HQSS. However, in this case HQSS should not be satisfied, because the heavy quark propagator goes as $(1/m_{D^*_s})^2$ and those terms are subleading in the $(1/m_Q)$ counting ($m_Q$ is the mass of the heavy quarks).

\section{Baryon wave functions}
\label{sec:Baryon}

We need the baryon states of $J^P={1/2}^+$:
\begin{enumerate}

\item $\Xi^+_c$: $\frac{1}{\sqrt{2}} \,c	\,(us - su)$, and the spin wave function will be the mixed antisymmetric, $\chi_{MA}$,
for the two light quarks. Thus, the spin reads $\chi_c \frac{1}{\sqrt{2}}(\uparrow \downarrow - \downarrow \uparrow)$, with
$\chi_c =\, \uparrow \, \textrm{or}\, \downarrow$ for $S_z=1/2$ or $-1/2$;

\item $\Xi^0_c$: the same as $\Xi^+_c$, changing $(us - su) \to (ds - sd)$;

\item $\Xi^{\prime\,+}_c$: $\frac{1}{\sqrt{2}} c (us + su)$, and now the spin wave function for the three quarks
will be the mixed symmetric, $\chi_{MS}$, in the last two quarks

\begin{equation}\label{ms}
\chi_{MS}=
\left\{
  \begin{array}{rl}
    \hbox{$\displaystyle \frac{1}{\sqrt{6}}(\uparrow\uparrow\downarrow + \uparrow\downarrow\uparrow - 2 \downarrow\uparrow\uparrow)$}, & \hbox{for $S_z=1/2$}, \\
    \hbox{$\displaystyle -\frac{1}{\sqrt{6}}(\downarrow\uparrow\downarrow + \downarrow\downarrow\uparrow - 2 \uparrow\downarrow\downarrow)$}, & \hbox{for $S_z=-1/2$};
  \end{array}
\right.
\end{equation}

\item $\Xi^{\prime\,0}_c$: the same as $\Xi^{\prime}_c$, changing $(us+su) \to (ds+sd)$;

\item $\Omega^0_c$: $css$, and the spin wave function $\chi_{MS}$ in the last two quarks, like that for $\Xi^{\prime}_c$ ;

\item $\Xi^0$: to be consistent with the chiral Lagrangians one has to use a different phase convention with respect
to Ref.~\cite{close}, where the $\Sigma^+$, $\Xi^0$ and $\Lambda$ change sign with respect to Ref.~\cite{close}. The correct
assignment for the $\phi_{MA}$ are given in Table III of Ref.~\cite{miyahara} (the same assignment is also used in Ref.~\cite{Pavao:2017cpt}). Thus
\begin{align}
\label{xiwave}
\Xi^0 \equiv \frac{1}{\sqrt{2}}(\phi_{MS}\,\chi_{MS}+\phi_{MA}\,\chi_{MA}),
\end{align}
with
\begin{align}
\phi_{MS} = \frac{1}{\sqrt{6}} [s(us+su) - 2uss]\, ,
\end{align}
\begin{align}
\phi_{MA} = -\frac{1}{\sqrt{2}}[s(us - su)]\, ,
\end{align}
and $\chi_{MS}$ is given in Eq.~\eqref{ms}, while $\chi_{MA}$ is given by
\begin{equation}\label{ma}
\chi_{MA}=
\left\{
  \begin{array}{rl}
    \hbox{$\displaystyle \frac{1}{\sqrt{2}}\uparrow (\uparrow\downarrow - \downarrow\uparrow)$}, & \hbox{for $S_z=1/2$}, \\
    \hbox{$\displaystyle \frac{1}{\sqrt{2}}\downarrow (\uparrow\downarrow - \downarrow\uparrow)$}, & \hbox{for $S_z=-1/2$};
  \end{array}
\right.
\end{equation}

\item $\Xi^-$: as in Eq.~\eqref{xiwave} with
\begin{align}
\phi_{MS} = -\frac{1}{\sqrt{6}} [s(ds + sd) - 2dss] \, ,
\end{align}
\begin{align}
\phi_{MA} = \frac{1}{\sqrt{2}}[s(ds - sd)]\, .
\end{align}
For the baryon states of spin $J^P={3/2}^+$ we have

\item $\Xi^{*+}_c$: $c\frac{1}{\sqrt{2}}(us+su)$, and the symmetric spin
wave function, $\chi_S = \, \uparrow\uparrow\uparrow$, ... ;

\item $\Xi^{*0}_c$: $c\frac{1}{\sqrt{2}}(ds+sd)$, and $\chi_S$;

\item $\Omega^*_c$: $css$, and $\chi_S$;

\item $\Xi^{*0}$: $\frac{1}{\sqrt{3}}(sus + ssu +uss)$, and $\chi_S$;

\item $\Xi^{*-}$: $\frac{1}{\sqrt{3}}(sds + ssd +dss)$, and $\chi_S$;
\end{enumerate}

We have to construct states with $I=0$ to match the $\Omega_c$. For that recall that our isospin multiplets are:
\begin{eqnarray}
\bar{K} = \myvec{\bar{K}^0\\-K^-};\,\,D = \myvec{D^+\\-D^0};\,\,
 \Xi = \myvec{\Xi^0\\-\Xi^-};\,\, \Xi^\ast = \myvec{\Xi^{\ast 0}\\ \Xi{^{\ast -}}}; \nonumber\\
 \vspace{15pt}
 \Xi_c = \myvec{\;\Xi_c^+\\ \;\Xi_c^0};\,\,
 \Xi^{\prime}_c = \myvec{\;\Xi^{\prime +}_c\\ \; \Xi^{\prime 0}_c};\,\,\,\,\,
\,\,  \Xi^{*}_c = \myvec{\Xi^{* +}_c\\\Xi^{*0}_c}\, ;
\end{eqnarray}
and thus
\begin{eqnarray}
| \Xi_c \bar{K}, I=0 \rangle &=& -\frac{1}{\sqrt{2}} \Big|\Xi_c^+ K^- + \Xi_c^0 \bar{K}^0\Big\rangle\, ,\nonumber\\
| \Xi D, I=0 \rangle &=& -\frac{1}{\sqrt{2}} \Big|\Xi^0D^0 - \Xi^-D^+\Big\rangle\, ,\nonumber\\
| \Xi^*_c \bar{K}, I=0 \rangle &=& -\frac{1}{\sqrt{2}} \Big| \Xi^{* +}_c K^- + \Xi^{* 0}_c \bar{K}^0\Big\rangle\, ,\nonumber\\
| \Xi^* D, I=0 \rangle &=& -\frac{1}{\sqrt{2}} \Big|\Xi^{*0} D^0 + \Xi^{*-} D^+\Big\rangle\, .
\end{eqnarray}
With these wave functions and the prescription to calculate the $VPP$ and $VBB$ vertices we can construct the matrix elements of the transition potential between the states of Table \ref{tab1}. Some examples are shown in Appendix \ref{app}.

Following the steps of Appendix \ref{app} it becomes easy and systematic to evaluate all the matrix elements and we find
\begin{equation}
\label{vij}
V_{ij}= D_{ij}\frac{1}{4f_\pi^2} (p^0+ p^{\prime 0}) \, .
\end{equation}

Alternatively, we can use another expression which includes relativistic correction in $s$-wave \cite{Bennhold}
\begin{equation}
    \label{vijRelat}
V_{ij}= D_{ij}\frac{2\sqrt{s} -M_{B_i} -M_{B_j}}{4f_\pi^2}  \sqrt{\frac{M_{B_i} + E_{B_i}}{2M_{B_i}}} \sqrt{\frac{M_{B_j} + E_{B_j}}{2M_{B_j}}}\, ,
\end{equation}
where $M_{{B_i},{B_j}}$ and $E_{{B_i},{B_j}}$ stand for the mass and the center-of-mass energy of the baryons, respectively, and the matrix $D_{ij}$ is given in Table \ref{tabDij}.

\begin{table}[h!]
\label{tabDij}
\caption{$D_{ij}$ coefficients of Eq.~\eqref{vijRelat} for the meson-baryon states coupling to $J^P={1/2}^-$ in $s$-wave.}
\centering
\begin{tabular}{c || c c c c c c c}
\hline\hline
 $J=1/2$~ & ~~$\Xi_c\bar{K}$~ & ~$\Xi^{\prime}_c\bar{K}$~ & ~$\Xi D$~ & ~$\Omega_c \eta$ ~&~ $\Xi D^*$ ~& ~$\Xi_c \bar{K}^*$ ~& ~$\Xi^{\prime}_c\bar{K}^*$\\
\hline\hline
$\Xi_c\bar{K}$ & $-1$ & $0$ & $-\frac{1}{\sqrt{2}}\lambda$ & $0$ & $0$ & $0$ & $0$\\
$\Xi^{\prime}_c\bar{K}$ &  & $-1$ & $\frac{1}{\sqrt{6}}\lambda$ & $-\frac{4}{\sqrt{3}}$ & $0$ & $0$ & $0$\\
$\Xi D$ & & & $-2$ & $\frac{\sqrt{2}}{3}\lambda$ & $0$ & $0$ & $0$\\
$\Omega_c \eta$ & & &  & $0$ & $0$ & $0$ & $0$\\
$\Xi D^*$ & & & & & $-2$ & $-\frac{1}{\sqrt{2}}\lambda$ & $\frac{1}{\sqrt{6}}\lambda$\\
$\Xi_c \bar{K}^*$ & & & & & & $-1$ & $0$\\
$\Xi^{\prime}_c\bar{K}^*$ & & & & & & & $-1$\\
\hline\hline
\end{tabular}
\end{table}

In Table \ref{tabDij} we have the parameter $\lambda$ in some non diagonal matrix elements, which involve transitions from one meson without charm to one with charm, like $\bar{K} \to D$. In this case we have for the propagator of the exchanged vector
\begin{eqnarray}
\frac{1}{(q^0)^2 - |{\bf q\,}|^2-m_{D^*_s}^2} \approx \frac{1}{(m_D - m_K)^2-m_{D^*_s}^2}\, ,
\end{eqnarray}
and the ratio to the propagator of the light vectors is
\begin{eqnarray}
\lambda \equiv \frac{-m^2_{V}}{(m_D - m_K)^2 - m^2_{D^*_s}}\approx 0.25 \, .
\end{eqnarray}
We take $\lambda = 1/4$ in all these matrix elements, as it was done in Ref.~\cite{mizutani}.

The diagonal matrix elements of Table \ref{tabDij} coincide with those of Ref.~\cite{angels},
but not all the non diagonal. This is not surprising. $\textrm{SU(4)}$ symmetry is used in Ref.~\cite{angels},
but only $\textrm{SU(3)}$ is effectively used in the diagonal terms, as we have argued. Then we should note that
the heavy baryons that we have constructed are not eigenstates of $\textrm{SU(4)}$ since we have singled out the heavy quarks
and used symmetrized wave functions for the light quarks. This induces a spin-flavor dependence different from the
one of pure $\textrm{SU(4)}$ symmetry.

With respect to Ref.~\cite{romanets}, we have some diagonal matrix elements equal but not all of them, and there are also differences
in the non diagonal terms. These matrix elements are also different from those of Ref.~\cite{angels}.

\begin{table}[h!]
\caption{$D_{ij}$ coefficients of Eq.~\eqref{vijRelat} for the meson-baryon states coupling to $J^P={3/2}^-$.}
\centering
\begin{tabular}{c || c c c c c c}
\hline\hline
$J=3/2$ ~~ & ~~$\Xi^*_c \bar{K}$ ~&~ $\Omega^*_c \eta$ ~&~ $\Xi D^*$ ~&~ $\Xi_c \bar{K}^*$ ~&~ $\Xi^* D$ ~&~ $\Xi^{\prime}_c \bar{K}^*$\\
\hline\hline
$\Xi^*_c \bar{K}$ & $-1$ & $-\frac{4}{\sqrt{3}}$ & $0$ & $0$ & $\frac{2}{\sqrt{6}}\lambda$ & $0$\\
$\Omega^*_c \eta$ & & $0$ & $0$ & $0$ & $-\frac{\sqrt{2}}{3}\lambda$ & $0$\\
$\Xi D^*$ & & & $-2$ & $-\frac{1}{\sqrt{2}}\lambda$ & $0$ & $\frac{1}{\sqrt{6}}\lambda$\\
$\Xi_c \bar{K}^*$ & & & & $-1$ & $0$ & $0$\\
$\Xi^* D$ & & & & & $-2$ & $0$\\
$\Xi^{\prime}_c \bar{K}^*$ & & & & & & $-1$\\
\hline\hline
\end{tabular}
\label{tabDij2}
\end{table}

To calculate the matrix elements for the states that couple to $J^P={3/2}^-$ of Table \ref{tab2} we proceed in the same way as in Appendix \ref{app}. We must take into account that the $VVV_{ex}$ are like those of $PPV_{ex}$ under the approximation of neglecting $({\bf p}/m_V)^2$, where ${\bf p}$ is the momentum of the external vector. In addition one has for the factor $\vec{{\bf \epsilon}} \cdot  \vec{{\bf \epsilon\,}}^{\prime}$ for the vector polarization, which makes these terms to contribute to $J=1/2$ and $J=3/2$ with degeneracy. The terms connecting $P$ and $V$ like $\Xi^*_c \bar{K} \to \Xi D^*$ require exchange of pseudoscalars, which go with the momentum and are small compared to the exchange of vectors \cite{xiaojuan}. In the $\Xi^*_c \bar{K} \to \Xi D^*$ one would have to exchange a $D_s$ and it would be doubly suppressed. In the $\Xi^*_c \bar{K} \to \Xi K^*$ one would exchange a pion, but the $K$ and $K^*$ states are quite separated in energy and the transition is also not important. In the $D \Xi \to D^* \Xi$ transitions one has the $\pi \Xi \Xi$ Yukawa vertex that goes like $D-F$ compared to $D+F$ for $\pi PP$, with $F=0.51$, $D=0.75$ \cite{Borasoy}, which is highly suppressed. Therefore, we neglect all terms which involve transition of a pseudoscalar to a vector and then the matrix elements are again given by Eq.~\eqref{vijRelat} with the $D_{ij}$ coefficients given in Table \ref{tabDij2}.

\begin{figure}
  \centering
    \includegraphics[width=\textwidth]{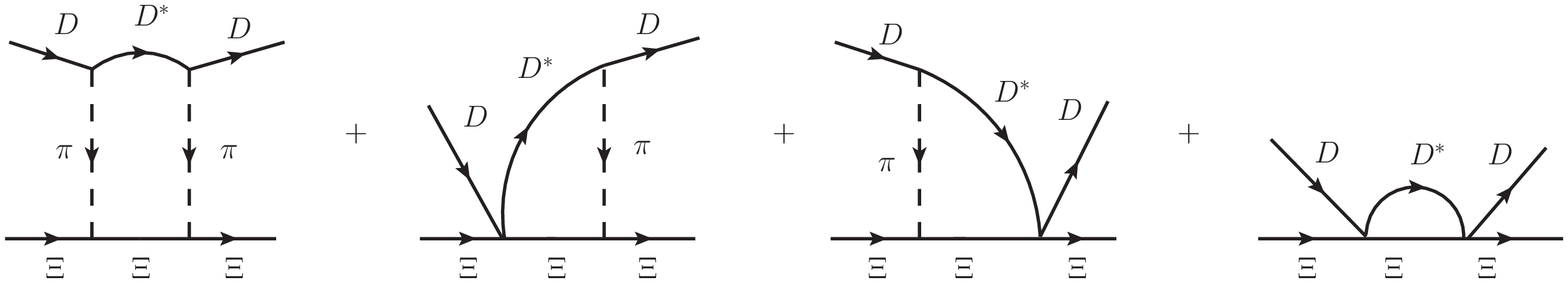}\\
  \caption{Box diagrams accounting for the $D \, \Xi \to D^* \, \Xi \to D \, \Xi$.}\label{pionX}
\end{figure}

In order to see the relevance of the $\pi$ exchange discussed above, we take the $D \, \Xi \to D^* \, \Xi$ transition and we evaluate the effect in the $D \, \Xi \to D \, \Xi$ interaction going through the intermediate $D^* \, \Xi$ state. For this we follow Ref.~\cite{xiao} and consider the diagrams of Fig.~\ref{pionX}.

As discussed in Ref.~\cite{xiao}, in addition to the $\pi$ exchange there is a contact term called Kroll-Ruderman in the $\gamma \, N \to \pi \, N$ (or $\rho \, N \to \pi \, N$) transition, then the four diagrams of Fig.~\ref{pionX} must be evaluated. They provide a $\delta V$ potential for $D \, \Xi \to D \, \Xi$ which can be evaluated by means of Eq.~(40) of Ref.~\cite{xiao}, simply changing the masses of $B$, $B^*$ to $D$, $D^*$ and $N$ to $\Xi$. We have performed the calculation and, compared to the potential $V_{ij}$ from Eq.~\eqref{vijRelat} and Table \ref{tabDij} we find $\delta V / V \approx 0.012$ for the $\Xi \, D$ channel calculated at the energy of the pole around 3090 MeV (which is dominated by this channel, as shown in Table \ref{res1} of the Results section), a correction of order $1\%$, which we safely neglect.

\section{Results}
\label{sec:res}

We use the potential of Eq.~\eqref{vijRelat} and the on-shell factorized Bethe-Salpeter equation \cite{nsd,ollerulf}
\begin{equation}
\label{bs}
T = [1 - V G]^{-1}\, V \, ,
\end{equation}
 with $G$ the meson-baryon loop function. We choose to regularize it with the cutoff method to
 avoid potential pathologies of the dimensional regularization in the charm sector, where $G$ can
 become positive below threshold (and eventually produce bound states with a repulsive potential) \cite{zouwu}.
 There is another reason, because in order to respect rules of heavy quark symmetry in bound states, it was shown
 in Refs.~\cite{lugeng,xiaooz} that the same cutoff has to be used in all cases. Alternatively one can use a special
 $G$ function defined in Ref.~\cite{altenbu}.

 The $G$ function for meson-baryon with the cutoff method is given by
 \begin{eqnarray}
 G_l &=& i \int \frac{d^4q}{(2\pi)^4} \frac{M_l}{E_l({\bf q})}\frac{1}{k^0+p^0 - q^0 - E_l({\bf q})+i\epsilon} \frac{1}{{\bf q}^2-m^2_l+i\epsilon}\nonumber\\
 &=& \int_{|{\bf q}|< q_{max}} \frac{d^3q}{(2\pi)^3}\frac{1}{2\omega_l({\bf q})}\frac{M_l}{E_l({\bf q})}\frac{1}{k^0+p^0-\omega_l({\bf q}) - E_l({\bf q})+i\epsilon}\, ,
 \end{eqnarray}
where $k^0+p^0=\sqrt{s}$ and $\omega_l$, $E_l$, are the energies of the meson and baryon respectively and $m_l,\,M_l$ the meson and baryon masses.

We evaluate Eq.~\eqref{bs} and look for poles in the second Riemann sheet, where we define $G^{II}_l$ for
${\rm Re}(\sqrt{s})$ bigger than the threshold of the $l$ channel as
\begin{eqnarray}
G^{II}_l = G^I_l + i \frac{2M_l\,q}{4\pi\sqrt{s}}\, ,
\end{eqnarray}
with $q$ given by
\begin{eqnarray}
q = \frac{\lambda^{1/2}(s,m^2_l,M^2_l)}{2\sqrt{s}}\, ,\, \textrm{and Im(q)}>0\, .
\end{eqnarray}

In addition, we evaluate the couplings $g_i$ of the states obtained to the different channels defined such that, close to the pole $M_R$ we have
\begin{equation}
T_{ij} = \frac{g_i g_j}{\sqrt{s}-M_R} \, ,
\end{equation}
and we also evaluate $g_iG_i$, which for $s$-wave gives the strength of the wave function at the origin \cite{gamermann}.

In Table \ref{res2} we show the poles that we obtain for the $J^P={1/2}^-$ sector for different values of the cutoff $q_{max}$. We only show the results with the pseudoscalar-baryon interaction. This sector decouples from the vector-baryon one, where the states are obtained degenerate in $J^P=1/2^-, 3/2^-$ . We will come back to this sector later on.

\begin{table}[h!]
\caption{Poles in $J^P={1/2}^-$ sector from pseudoscalar-baryon interaction. (Units: MeV).}
\centering
\begin{tabular}{| c | c | c | c | c | c |}
\hline\hline
$q_{max}$ & 600 & 650 & 700 & 750 & 800\\
\hline
  & $3065.4+i0.1$ & ${\bf 3054.05+i0.44}$ & $3038.13+i1.78$ & $3016.21+i6.02$ & $2989.69+i16.24$\\
& $3114.22+i3.75$ &  ${\bf 3091.28+i5.12}$ & $3067.71+i4.12$ & $3046.24+i3.83$ & $3027.75+i2.19$\\
\hline\hline
\end{tabular}
\label{res2}
\end{table}


We can see that we always get two states in the range of the masses observed experimentally. The
strategy followed in these calculations is to fine tune the cutoff to adjust the pole position to some experimental data. We see that if we take $q_{max}=650$ MeV the results agree well with the second and fourth resonances reported in Ref.~\cite{lhcbexp}, $\Omega_c(3050)$ and $\Omega_c(3090)$. It is interesting to note that cutoff values of this order are used in Ref.~\cite{ramoskaon} for $\bar{K}N$ or in Ref.~\cite{uchino} for $DN$. Fitting one resonance is partly merit of fine tuning the cutoff, but then the second resonance and the widths are genuine predictions of the theory. Note that the widths are respectively $0.88$ MeV and $10.24$ MeV  which agree remarkably well with the experiment, $0.8\pm 0.2\pm 0.1$ MeV and $8.7\pm 1.0 \pm 0.8$ MeV, respectively. It is instructive to see the origin of the widths. For this we look at Table \ref{res1} for the couplings to the different channels. We can see that for the lower state at $3054$ MeV only the $\Xi_c\bar{K}$ state is open for decay, precisely the channel where it has been observed, and the coupling of the state to this channel is very small. However, for the state at $3091$ MeV the $\Xi^{\prime}_c\bar{K}$ channel is also open, and the coupling to this channel is considerable. Furthermore, the coupling to $\Xi_c\bar{K}$ is bigger than before and there is more phase space for decay.

\begin{table}[h!]
\caption{The coupling constants to various channels for the poles in the $J^P={1/2}^-$ sector, with $q_{max}=650$ MeV, and $g_i\,G^{II}_i$ in MeV.}
\centering
\begin{tabular}{c c c c c c c c}
\hline\hline
${\bf 3054.05+i0.44}$ & $\Xi_c\bar{K}$ & $\Xi^{\prime}_c\bar{K}$ & $\Xi D$ & $\Omega_c \eta$ & $\Xi D^*$ & $\Xi_c \bar{K}^*$ & $\Xi^{\prime}_c\bar{K}^*$\\
\hline
$g_i$ & $-0.06+i0.14$ & $\bf 1.94+i0.01$ & $-2.14+i0.26$ & $1.98+i0.01$ & 0 & 0 & 0\\
$g_i\,G^{II}_i$ & $-1.40-i3.85$ & $\bf -34.41-i0.30$ & $9.33-i1.10$ & $-16.81-i0.11$ & 0 & 0 & 0\\
\hline\hline
${\bf 3091.28+i5.12}$ & $\Xi_c\bar{K}$ & $\Xi^{\prime}_c\bar{K}$ & $\Xi D$ & $\Omega_c \eta$ & $\Xi D^*$ & $\Xi_c \bar{K}^*$ & $\Xi^{\prime}_c\bar{K}^*$\\
\hline
$g_i$ & $0.18-i0.37$ & $0.31+i0.25$ & $\bf 5.83-i0.20$ & $0.38+i0.23$ & 0 & 0 & 0\\
$g_i\,G^{II}_i$ & $5.05+i10.19$ & $-9.97-i3.67$ & $\bf -29.82+i0.31$ & $-3.59-i2.23$ & 0 & 0 & 0\\
\hline\hline
\end{tabular}
\label{res1}
\end{table}

Next we look for the states of $J^P={3/2}^-$ from the pseudoscalar-baryon$({3/2}^+)$ interaction. In Table \ref{tabDij2} we see that the pseudoscalar-baryon$({3/2}^+)$ states do not
couple to vector-baryon and we can separate two blocks, the channels $\Xi^*_c\,\bar{K}$, $\Omega_c^*\,\eta$, $\Xi^*\,D$ and
$\Xi\,D^*$, $\Xi_c\,\bar{K}^*$, $\Xi^{\prime}_c\,\bar{K}^*$. The first three channels in $s$-wave give rise to $J^P={3/2}^-$, while
the other three give rise to $J^P={1/2}^-, {3/2}^-$, degenerated in our approach. We then separate these two
sets of states.

In Table \ref{res3} we show the results for $J^P={3/2}^-$ for different values of the cutoff.
We see that we get two poles. Yet, if we choose the same cutoff as in the $J^P={1/2}^-$ sector we find a mass of
$3125$ MeV and zero width for the lowest state. As we can see, the mass is smaller than all the thresholds in Table~\ref{tab2}, hence it does not decay into them. To decay into $\Xi_c\bar{K}$, where it has been observed, we would need the exchange of vector mesons in $p$-wave, which give rise to a small width. We can clearly associate the state found with the $\Omega_c(3119)$ observed experimentally, which has a width of $1.1\pm0.8\pm0.4$ MeV. The agreement is also remarkable.

\begin{table}[h!]
\caption{Poles in $J^P={3/2}^-$ sector from pseudoscalar-baryon($3/2^+$) interaction. (Units: MeV).}
\centering
\begin{tabular}{| c | c | c | c | c | c | c | c | c | c |}
\hline\hline
$q_{max}$ & 600 & 650 & 700 & 750 & 800 \\
\hline
 \,& \,$3134.39$\, & \,${\bf 3124.84}$ \,& \,$3112.83$ \,& \,$3099.2$ \,& \,$3084.52$\, \\
 \,& \,$3316.48+i0.14$ \,&\, $3290.31+i0.03$\,& \,$3260.42+i0.08$ \,& \,$3227.34+i0.15$ \,& \,$3191.13+i0.22$\,\\
\hline\hline
\end{tabular}
\label{res3}
\end{table}

In Table \ref{res4} we show the couplings of the states to the coupled channels of Table~\ref{tab2}. We can see that
the state at $3125$ MeV couples strongly to $\Xi_c^*\,\bar{K}$ and $\Omega_c^*\,\eta$, more strongly
to $\Xi_c^*\,\bar{K}$. The upper state couples very strongly to $\Xi^*\,D$.

\begin{table}[h!]
\caption{The coupling constants to various channels for the poles in the $J^P={3/2}^-$ sector, with $q_{max}=650$ MeV, and $g_i\,G^{II}_i$ in MeV.}
\centering
\begin{tabular}{c c c c c c c}
\hline\hline
${\bf 3124.84}$ & $\Xi^*_c \bar{K}$ ~&~ $\Omega^*_c \eta$ ~&~ $\Xi D^*$ ~&~ $\Xi_c \bar{K}^*$ ~&~ $\Xi^* D$ ~&~ $\Xi^{\prime}_c \bar{K}^*$\\
\hline
$g_i$ & $\bf 1.95$ & $1.98$ & $0$ & $0$ & $-0.65$ & $0$\\
$g_i\,G^{II}_i$ & $\bf -35.65$ & $-16.83$ & $0$ & $0$ & $1.93$ & $0$\\
\hline\hline
$3290.31+i0.03$ & $\Xi^*_c \bar{K}$ ~&~ $\Omega^*_c \eta$ ~&~ $\Xi D^*$ ~&~ $\Xi_c \bar{K}^*$ ~&~ $\Xi^* D$ ~&~ $\Xi^{\prime}_c \bar{K}^*$\\
\hline
$g_i$ & $0.01+i0.02$ & $0.31+i0.01$ & $0$ & $0$ & $\bf 6.22-i0.04$ & $0$\\
$g_i\,G^{II}_i$ & $-0.62-i0.18$ & $-5.25-i0.18$ & $0$ & $0$ & $\bf -31.08+i0.20$ & $0$\\
\hline\hline
\end{tabular}
\label{res4}
\end{table}

For the vector-baryon states with $J^P={1/2}^-, {3/2}^-$ we choose the same cutoff $q_{max}=650$ MeV that we
have chosen in the former cases and find three states that we show in Table \ref{res5}
together with the couplings to each channel.

\begin{table}
\caption{The coupling constants to various channels for the poles  in $J^P={1/2}^-, {3/2}^-$ stemming from vector-baryon interaction with $q_{max}=650$ MeV, and $g_i\,G^{II}_i$ in MeV.}
\centering
\begin{tabular}{c c c c}
\hline\hline
$3221.98$ & $\Xi\,D^*$ & $\Xi_c\,\bar{K}^*$ & $\Xi^{\prime}_c\,\bar{K}^*$\\
\hline
$g_i$ & $\bf 6.37$ & $0.59$ & $-0.28$ \\
$g_i\,G^{II}_i$ & $\bf -29.29$ & $-4.66$ & $1.62$\\
\hline\hline
$3360.37+i0.20$ \,&\, $\Xi\,D^*$ \,&\, $\Xi_c\,\bar{K}^*$ \,&\, $\Xi^{\prime}_c\,\bar{K}^*$\\
\hline
$g_i$ \,&\, $-0.11 -i0.12$ \,&\, $\bf 1.31 -i0.03$ \,&\, $0.03 +i0.01$\\
$g_i\,G^{II}_i$ \,&\, $2.12 +i0.48$ \,&\, $\bf -26.04 +i0.36$ \,&\, $-0.26 -i0.06$\\
\hline\hline
$3465.17+i0.09$ \,&\, $\Xi\,D^*$ \,&\, $\Xi_c\,\bar{K}^*$ \,&\, $\Xi^{\prime}_c\,\bar{K}^*$\\
\hline
$g_i$ \,&\, $-0.01+i0.06$ \,&\, $0.01-i0.01$ \,&\, $\bf 1.75+i0.01$\\
$g_i\,G^{II}_i$ \,&\, $-0.84-i0.23$ \,&\, $0.17+i0.24$ \,&\, $\bf -32.29-i0.08$\\
\hline\hline
\end{tabular}
\label{res5}
\end{table}

The first state obtained has zero width and couples mostly to $\Xi\,D^*$ while the second and third ones have very small widths and couple mostly to $\Xi_c\,\bar{K}^*$ and $\Xi^{\prime}_c\,\bar{K}^*$, respectively.
The widths could be bigger if we had considered vector-baryon transitions to pseudoscalar-baryon but we argued that they were small in any case and neglected them in our study.

It is interesting to compare our results with those of  Ref.~\cite{angels}. The main feature is that the results obtained are remarkably similar. In Ref.~\cite{angels} two states of $J^P={1/2}^-$ are also found that compare well with the $\Omega_c(3050)$ and $\Omega_c(3090)$, as we have found here.
The width of the second state is about $17$ MeV, while we get $10$ MeV, closer to the experimental value. In Ref.~\cite{angels} two sets of subtraction constants (cutoffs) are used and in one of them the width of this state is $12$ MeV, at the expense of using a somewhat small cutoff in the $\Xi_c\,\bar{K}$ decay channel of $320$ MeV. Even then, the main channels and the strengths of the couplings are similar to ours.

In Ref.~\cite{angels} the compositeness magnitude $-g^2\,\partial G/\partial\sqrt{s}$ is evaluated for all channels.
This magnitude provides the probability to find bound channels \cite{gamermann,sekihara,kamiya}
and for the case of open channels it gives the integral of the wave functions squared with a given prescription of the phase \cite{acetifun}. The magnitude $g\,G$ that we calculate gives the strength of each channel to produce the resonance (up to coefficients appearing in the primary steps of a reaction prior to final state interaction). Yet, there is a correspondence in these two magnitudes, and we find that when $-g^2\,\partial G/\partial\sqrt{s}$ is large for some channel in Ref.~\cite{angels}, so is $g\,G$ in our case.

The pseudoscalar-baryon(${3/2}^+$) states are not considered in Ref.~\cite{angels} and, thus, the states that we get in Table \ref{res3} are new. As to the vector-baryon(${1/2}^+$) states we obtain three new states, two of them in qualitative agreement with Ref.~\cite{angels}. In Ref.~\cite{angels} two states were found at $3231$ MeV and $3419$ MeV, that couple mostly to $\Xi\,D^*$ and $\Xi^{\prime}_c\,\bar{K}^*$, respectively. We also find two states, at
$3222$ MeV and $3465$ MeV, which also couple mostly to $\Xi\,D^*$ and $\Xi^{\prime}_c\,\bar{K}^*$, respectively, as in Ref.~\cite{angels}, plus a new intermediate state at 3360 MeV that couples mostly to $\Xi_c\,\bar{K}^*$.

As to the results of Ref.~\cite{romanets}, the bindings obtained there, in the absence of any experimental data, gave rise to
bound $\Omega_c$ states with more binding than here. It would be interesting to have a new look in that
framework under the light of the new experimental information.

The basic input of our calculations are the $V_{ij}$ transition potentials of Eq.~\eqref{vijRelat}, and the coupling that we have is $\displaystyle \frac{1}{f_\pi^2}$. We estimate uncertainties in the following way. We increase $f_\pi^2$ by $10\%$ and readjust the cutoff to obtain the same energy of the first state (going from $q_{\rm max} = 650$ MeV to $694$ MeV), and then we get the results of Table \ref{tab:Uncert}.
As we can see, the changes in the masses and widths are small. The difference in the masses are always smaller than 5 MeV, and for the three states that we compare with experiment the changes are even smaller. The widths also change a bit, but the width of the widest state only changes from 10.24 MeV to 11.82 MeV, and the others are still very small and compatible with experiment within errors.

\begin{table}
\caption{Dependence of the results on the value of $f_\pi$.}
\centering
\begin{tabular}{c | c | c}
\hline\hline
$J=1/2$ & $\,f_\pi = 93$ MeV and $q_{\rm max} = 650$ MeV & $\,f_\pi = 97.6$ MeV and $q_{\rm max} = 694$ MeV \\
\hline
Pole 1 & $\bf 3054.05+i0.44$ & $3054.05+i0.70$ \\
Pole 2 & $\bf 3091.28+i5.12$ & $3087.24+i5.91$ \\
\hline\hline
$J=3/2$ & $\,f_\pi = 93$ MeV and $q_{\rm max} = 650$ MeV & $\,f_\pi = 97.6$ MeV and $q_{\rm max} = 694$ MeV \\
\hline
Pole 1 & $\bf 3124.84$ & $3125.71$ \\
Pole 2 & $3290.31+i0.03$ & $3284.73.24+i0.05$ \\
\hline\hline
$J=1/2, \, 3/2\,$ & $\,f_\pi = 93$ MeV and $q_{\rm max} = 650$ MeV & $\,f_\pi = 97.6$ MeV and $q_{\rm max} = 694$ MeV \\
\hline
Pole 1 & $3221.98$ & $3216.98$ \\
Pole 2 & $3360.37+i0.20$ & $3361.28 +i0.18$ \\
Pole 3 & $3465.17+i0.09$ & $3469.04 +i0.07$ \\
\hline\hline
\end{tabular}
\label{tab:Uncert}
\end{table}

\section{Conclusions}
\label{sec:conc}

We have studied $\Omega_c$ states which are dynamically generated from the interaction of meson-baryon
in the charm sector. The interaction is obtained using an extension of the local hidden gauge approach with the exchange of vector
mesons. We show that the dominant terms come from exchange of light vector mesons, leaving the heavy quarks
as spectators. This has two good consequences: first we can map the interaction to what happens in $\textrm{SU(3)}$
using chiral Lagrangians, and second, the fact that the heavy quarks are spectators in the interaction guarantee that
the dominant terms in the $(1/m_Q)$ counting fulfill the rules of heavy quark symmetry.

We obtain two states with $J^P={1/2}^-$ which are remarkably close in mass and width to the experimental states
$\Omega_c(3050)$, $\Omega_c(3090)$. In addition, we also obtain a ${3/2}^-$ state with zero width at $3125$ MeV, which
can be associated to the experimental $\Omega_c(3119)$ that also has a width of the order or smaller than $1$ MeV.

The agreement of the results with experiment is remarkable. It would be very interesting to see the next experimental steps
to determine the spin-parity of these states, which could serve to discriminate between present models where there are
large discrepancies concerning the spin-parity assignment.

\vspace{-0.3cm}
\begin{acknowledgments}
V. R. Debastiani acknowledges the Programa Santiago Grisolia of Generalitat Valenciana (Exp. GRISOLIA/2015/005). J. M. Dias thanks the Funda\c c\~ao de Amparo \`a Pesquisa do Estado de S\~ao Paulo (FAPESP) for support by FAPESP grant 2016/22561-2. This work is partly supported by the Spanish Ministerio de Economia y Competitividad and European FEDER funds under the contract number 
FIS2014-57026-REDT, FIS2014-51948-C2- 1-P, and FIS2014-51948-C2-2-P, and the Generalitat Valenciana in the program Prometeo II-2014/068 (EO). This work is also partly supported by the National Natural Science Foundation of China under Grants No. 11565007, No. 11747307 and No. 11647309.
\end{acknowledgments}

\appendix
\section{Evaluation of the transition matrix elements of $\bar{K}\Xi_c\to \bar{K}\Xi_c $}
\label{app}

\begin{figure}[h!]
	\begin{center}
		\includegraphics[width=1.0\textwidth]{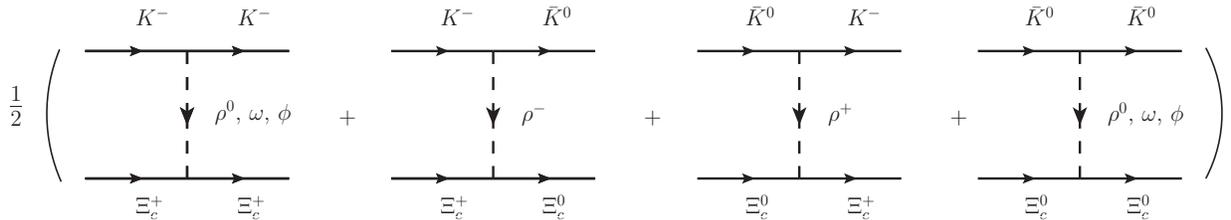}
	\end{center}
	\caption{\label{diag1App} Diagrams in the $\bar{K}\Xi_c\to \bar{K}\Xi_c $ transition.}
\end{figure}

We need to evaluate de diagrams of Fig.~\ref{diag1App}. The upper vertices are readily evaluated using Eq.~\eqref{vppLag}, and
we get
\begin{eqnarray}
-it_{K^-\to K^-} \myvec{\rho^0\\ \omega \\ \phi} &=& g V_{\mu}\,(-ip^{\mu} -i p^{\prime \mu})\myvec{1/\sqrt{2}\\ 1/\sqrt{2} \\ -1}\, ,\nonumber \\
-it_{K^-\to \bar{K}^0 \rho^-} &=& g \rho^{+\mu}\,(-ip^{\mu} -i p^{\prime \mu})\, ,
\end{eqnarray}
with $p,\,p^{\prime}$ the momenta of the incoming and outgoing kaons. We also have
\begin{eqnarray}
-it_{\bar{K}^0\to K^-\rho^+} &=& g \rho^{-\mu}\,(-ip^{\mu} -i p^{\prime \mu})\, ,\nonumber \\
-it_{K^0\to \bar{K}^0}\myvec{\rho^0\\ \omega \\ \phi}&=&g V_{\mu}\,(-ip^{\mu} -i p^{\prime \mu})\myvec{-1/\sqrt{2}\\ 1/\sqrt{2} \\ -1}\, .
\end{eqnarray}

\begin{figure}[h!]
	\begin{center}
		\includegraphics[width=0.8\textwidth]{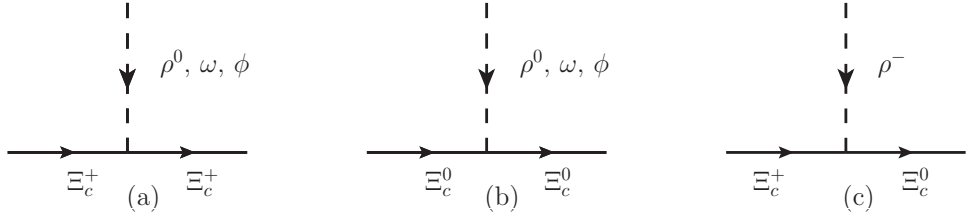}
	\end{center}
	\caption{\label{diag2App} Vector-baryon vertex diagrams in the $\bar{K}\Xi_c\to \bar{K}\Xi_c $ transition.}
\end{figure}

The lower vertices are readily calculated as seen in Fig.~\ref{diag2App}. 

For Fig.~\ref{diag2App}(a) we have the matrix elements

\begin{eqnarray}
\frac{1}{\sqrt{2}}\langle (us - su) | \myvec{g\frac{1}{\sqrt{2}}(u\bar{u}-d\bar{d})
\\g\frac{1}{\sqrt{2}}(u\bar{u}+d\bar{d})\\ g s\bar{s}}
 | \frac{1}{\sqrt{2}}(us-su)\rangle
= \myvec{\frac{1}{\sqrt{2}}\, g\\ \frac{1}{\sqrt{2}} \,g\\ g} \, .
\end{eqnarray}

For Fig.~\ref{diag2App}(b) we have
\begin{eqnarray}
\frac{1}{\sqrt{2}}\langle (ds - sd) | \myvec{g\frac{1}{\sqrt{2}}(u\bar{u}-d\bar{d})
\\g\frac{1}{\sqrt{2}}(u\bar{u}+d\bar{d})\\ g s\bar{s}}
 | \frac{1}{\sqrt{2}}(ds-sd)\rangle
= \myvec{-\frac{1}{\sqrt{2}}\, g\\ \frac{1}{\sqrt{2}} \,g\\ g}\, .
\end{eqnarray}

The vertices of Fig.~\ref{diag2App}(c) can be equally evaluated using the operator $g\,d\bar{u}$, or
simply one can use Clebsch-Gordan coefficients to relate to $\rho^0 \Xi_c^{+}\Xi_c^{0}$ and
we find the matrix element with the value $g$.

Altogether, the matrix element for Fig.~\ref{diag1App} is given by
\begin{eqnarray}
-it&=&\frac{1}{2} g^2 \Bigg{[}(-ip^{\mu} -i p^{\prime \mu})(-g_{\mu0})\frac{i}{-m^2_V}\myvec{1/\sqrt{2}\\ 1/\sqrt{2} \\ -1}
\,i\,\myvec{1/\sqrt{2}\\ 1/\sqrt{2} \\ 1} + 2 (-ip^{\mu} -i p^{\prime \mu})(-g_{\mu0})\frac{i}{-m^2_V} i\, \nonumber\\
&+& (-ip^{\mu} -i p^{\prime \mu})(-g_{\mu0})\frac{i}{-m^2_V}\myvec{-1/\sqrt{2}\\ 1/\sqrt{2} \\ -1} \,i\,
\myvec{-1/\sqrt{2}\\ 1/\sqrt{2} \\ 1} \Bigg{]}\, \nonumber\\
&=&-1\frac{1}{4f_\pi^2}(p^0+p^{\prime\, 0})\equiv D \frac{1}{4f_\pi^2}(p^0+p^{\prime\, 0})\, ,
\end{eqnarray}
with $D=-1$.

  \end{document}